\documentclass[twocolumn,aps,showpacs,amsmath,superscriptaddress]{revtex4-2}
\usepackage[latin9]{luainputenc}
\usepackage{mathtools}
\usepackage{amsmath}
\usepackage{amssymb}
\usepackage{graphicx}
\usepackage[symbol]{footmisc}

\makeatletter

\usepackage{dcolumn}
\usepackage{bm}
\usepackage{ifpdf}
\usepackage{lineno}\usepackage{amsfonts}
\usepackage{bbm}
\usepackage{mathrsfs}
\usepackage{upgreek}
\usepackage{epstopdf}
\usepackage{setspace}
\usepackage[matrix,frame,arrow]{xypic}
\usepackage{rotating}
\usepackage{float}
\usepackage{hyperref}
\usepackage{color}
\definecolor{med-blue}{RGB}{25,25,112}
\hypersetup{colorlinks, linkcolor={blue},citecolor={blue}, urlcolor={blue}}

\newcommand{\ket}[1]{\vert{#1}\rangle}

\newcommand{\mprod}[3]{\langle{#1}\vert{#2}\vert{#3}\rangle}

\makeatother

\begin{document}
	
\title{Vector detection of AC magnetic fields by Nitrogen-Vacancy centers of single orientation in diamond}
\author{Pooja Lamba}
\thanks{These authors contributed equally to this work.}
\affiliation{Department of Physics, Bennett University, Greater Noida 201310, India}
\author{Akshat Rana}
\thanks{These authors contributed equally to this work.}
\affiliation{Department of Physics, Bennett University, Greater Noida 201310, India}
\author{Sougata Halder}
\affiliation{Department of Physics, IIT Jodhpur, Jodhpur, India}
\author{Siddharth Dhomkar}
\affiliation{Department of Physics, IIT Madras, Chennai, India}
\affiliation{Center For Quantum Information, Communication And Computing, IIT Madras, Chennai, India}
\author{Dieter Suter}
\affiliation{Fakult{\"a}t Physik, Technische Universit{\"a}t Dortmund, D-44221 Dortmund, Germany}
\author{Rama K. Kamineni}
\email{koti.kamineni@gmail.com}
\affiliation{Department of Physics, Bennett University, Greater Noida 201310, India}

\date{\today}
\begin{abstract}
{Nitrogen-Vacancy (NV) centers in diamond have useful properties
for detecting both AC and DC magnetic fields with high sensitivity
at nano-scale resolution. Vector detection of AC magnetic fields can
be achieved by using NV centers having three different orientations. Here,
we propose a method to achieve this by using NV centers of single
orientation. In this method, a static magnetic field is applied perpendicular
to the NV axis, leading to strong mixing of the $m_{s}=-1$ and $1$
electron spin states. As a result, all three electron spin transitions
of the triplet ground state have non-zero dipole moments, with each
transition coupling to a single component of the magnetic field. This
can be used to measure both strength and orientation of the applied
AC field. To validate the technique, we perform a proof of principle experiment using a subset of  ensemble NV centers in diamond, all having the same orientation.
} 
\end{abstract}


\maketitle

\section{Introduction}

Sensitive detection of magnetic fields is an essential task in many
areas of science, technology, and medicine. Sensors based on different
techniques have been developed for this purpose \cite{BudkerRomalis2007,Vengalattore_BECmagnetometry2007,Waters_NMRSens1958,Fagaly_SQUIDSrev2006}.
Among these, sensors based on Nitrogen-Vacancy (NV) centers in diamond
have the advantages of room-temperature operation and high-sensitivity
detection with nano-scale resolution \cite{RevDegen,RevJacques,RevWalsworth}.
Additionally, NV centers are useful for detecting both AC and DC magnetic
fields \cite{JW2008Nat,Degen_Mag2008APL,Lukin_Mag2008Nat,Staudacher2013,Maletinsky_MWsens2015njp,DegenScience2017,JelezkoScience2017,Jingfu_RFsens2023}.

Sensitive detection of AC magnetic fields produced by spins and charges
at micro- and nano-scales have interesting applications in magnetic
resonance and condensed matter physics. Methods based on different
technologies have been developed for this purpose \cite{Clarke_RFSQUID1995,vanderWeide_RFtip1997,Lee_MWmicroscope2000,Boehi_MWimgAtoms2010,Boehi_MWimgVapor2012,Ockeloen_MW_BECmag2013,vanderSar_SpinWaves2015,Degen_CurrentImg2017}.
In many cases, it is desirable to measure not only the magnitude,
but also the orientation of the magnetic fields. Important examples
where vector detection of AC fields is important include magnetic
excitations and current distributions in materials \cite{Yacoby_RevNVCondMat2018}.
Most of the existing methods for detecting AC magnetic fields in the
microwave (MW) frequency range lack vector detection capabilities
or nano-scale resolution.

Small sensor size, long-coherence times, and ${\textrm C_{3V}}$ symmetry
of NV centers can be beneficial for vector detection of DC and AC magnetic fields. Vector detection of DC and slow time-varying fields by using NV centers have been thoroughly investigated \cite{Awschalom_VecMag2010,Fang_VecMag2018,Clevenson_VecMag2018,Walsworth_VecMag2018,Budker_VecMag2020,Hollenberg_VecMag2020,Weggler_VecMag2020,Du_VecMag2020}. Here, we focus on vector detection of fields with frequency greater than 1 MHz.
For small static magnetic fields, the quantization axis of NV centers
is pointed along the NV axis. The electron spin dipole moments, in
general, are perpendicular to the quantization axis and hence only
the components of the applied AC fields that are perpendicular to
the quantization axis interact with the spin transitions of the center.
Therefore, to achieve vector detection of AC fields using conventional
technique, at least, three NV centers having varied orientations are
essential. This has been demonstrated by using ensemble NV centers
\cite{DuVectorMW}. However, the requirement of multiple NV centers
with different orientations puts restriction on the sensor size and
limits its applicability for nano-scale imaging. Recently, vector
detection of AC magnetic fields by a single NV center has been demonstrated
by using a method known as "rotating-frame Rabi magnetometry" \cite{Cappellaro_VecAC2021}.
In this work, the components of the applied AC field have been measured
by tuning the Rabi frequencies of NV spin transitions to different
resonance conditions. However, this method requires that the Rabi
frequencies of the electron spin transitions are comparable to their
transition frequencies, which is a stringent prerequisite. A vector
detection scheme using level anti-crossings of an NV center coupled
to a first-shell $^{13}$C nuclear spin has been reported in Ref.
\cite{Koti2020_LAC}. The anti-crossings used in this work occur when the
energy level splitting due to Zeeman interaction of the NV electron spin is equal to the splitting
due to hyperfine interaction of first-shell $^{13}$C nuclear spin.
The disadvantage of this method is that it require a first-shell $^{13}$C
atom; probability of finding such a configuration is merely $3.3\%$.
Moreover, the bandwidth of the sensor is very narrow.

NV centers with transverse magnetic field have been previously used
for electric-field sensing \cite{WrachtrupESens}, suppression of
electron spin decoherence \cite{BajajDecoh,Koti2020_LAC}, and magnetic
field angle sensing \cite{Yacoby_AngleSens2021npj}.

Here, we propose and demonstrate a novel method for vector detection
of AC magnetic fields via NV centers of single orientation, which
exploits the energy level anti-crossing that occurs due to an application
of the static magnetic field perpendicular to the NV axis. Particularly,
for magnetic fields of strength up to few tens of mT, the eigenstates
of the electron spin of the center at this level anti-crossing can
be approximately written as $\ket{0}$, $\frac{\ket{-1}-\ket{1}}{\sqrt{2}}$,
and $\frac{\ket{-1}+\ket{1}}{\sqrt{2}}$, where $\ket{m_{S}}=\ket{0},\ket{\pm1}$
refer to the eigenstates of the electron spin operator $S_{z}$ and
$z$ is orientated along the symmetry axis of the NV center. All three
transitions between these levels, including the transition in the
$m_{s}=\pm1$ subspace, have strong dipole moments and their directions
are perpendicular to each other. Only the component of the applied
AC field that is oriented parallel to the dipole moment can excite
the corresponding transition. This can be used to measure both strength and orientation of the applied AC field. The amplitude of
the vector components of the AC field can be measured by on-resonantly
exciting each of the three transitions. The frequencies of these transitions
can be tuned to match the frequency of the AC field by varying the
strength of the static magnetic field.

This paper is arranged as follows - In section II, we describe the
method, and in section III, we demonstrate the method experimentally
by using an ensemble of identically oriented NV centers in diamond.
Finally, in section IV we discuss and conclude.

\section{Methodology}

\begin{figure}
\includegraphics[width=9cm]{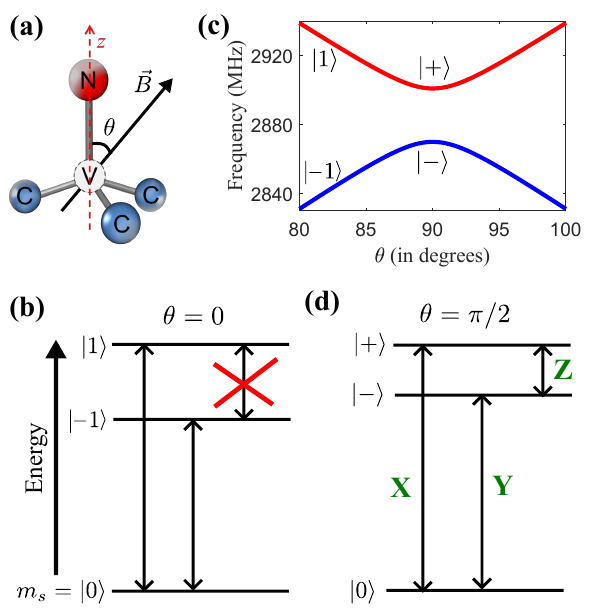} \caption{(a) Schematic diagram of the NV center in diamond and the static magnetic
field. (b) Energy level diagram of the ground state of the NV center
and the allowed transitions between them when $\vec{B}$ is oriented
parallel to the NV axis. (c) Frequencies of energy levels of the $m_{s}=\pm1$
subspace as a function of $\theta$ for $B=10.7$ mT. (d) Energy level
diagram when $\vec{B}$ is oriented perpendicular to the NV axis.
The transition labeled by Z has a dipole moment along the NV axis,
the transition labeled by X has a dipole moment along the direction
of $\vec{B}$, and the transition labeled by Y has a dipole moment
perpendicular to both these directions.}
\label{LACs}
\end{figure}

The NV center is a point defect in diamond with ${\textrm C_{3V}}$ symmetry
\cite{RevDoherty,SuterNVRev2017}. It has spin-1 ground and excited
states. Optical pumping causes polarization of its electron spin into
the $m_{S}=0$ state. Since the intensity of the fluorescence emitted
by the center depends on its spin state, it is possible to read the
spin state of the center optically. The Hamiltonian for the electron
spin of the ground state of the NV center in an applied static magnetic
field can be written as 
\begin{align}
{\cal H}_{0}= & DS_{z}^{2}+\gamma_{e}B(\sin\theta\ S_{x}+\cos\theta\ S_{z}).\label{eq:Hamiltonian}
\end{align}
Here, we consider $h=1$ and use frequency units for energy. $S_{x/y/z}$
represent the components of the spin-1 angular momentum operator.
$D=2870$ MHz is the zero-field splitting between the $\ket{m_{s}}=\ket{0}$
and $\ket{\pm1}$ states. $\gamma_{e}$ represent the gyromagnetic
ratio of the electron spin. $B$ and $\theta$ represent the strength
and polar angle of the static magnetic field in a coordinate system
whose Z-axis is aligned with the symmetry axis of the center, as shown
in Fig. \ref{LACs} (a). For $\theta\ne0$, the X-axis is defined
as the direction of the projection of the field into the plane perpendicular
to the Z-axis.

The $\ket{m_{s}}=\ket{\pm1}$ states, which are degenerate in zero
magnetic field, split due to the Zeeman interaction and this splitting
is proportional to the strength of the applied static magnetic field.
When the field is applied parallel to the NV axis or at a small angle
to it, $m_{s}$ is a good quantum number and the eigenstates of the
electron spin are $\ket{0}$, $\ket{-1}$, and $\ket{1}$. As shown
in Fig. \ref{LACs}(b), the transitions $\ket{0}\longleftrightarrow\ket{-1}$
and $\ket{0}\longleftrightarrow\ket{1}$ are allowed and the transition
between the states $\ket{-1}$ and $\ket{1}$ is forbidden. The transition
dipole moments of the allowed transitions are oriented perpendicular
to the NV axis and, thus, can be excited by components of resonant
microwave fields that are oriented perpendicular to the NV axis.

When the angle between the NV axis and the static magnetic field is
close to $\pi/2$, there is a strong mixing between the states $\ket{m_{s}}=\ket{-1}$
and $\ket{1}$. As shown in Fig. \ref{LACs}(c), the plot of energy
levels versus angle $\theta$ shows anti-crossing at $\theta=\pi/2$.
For magnetic fields of strength up to few tens of mT, the eigenstates
of the electron spin at $\theta=\pi/2$ can be approximately written
as $\ket{0}$, $\ket{-}$, and $\ket{+}$, where $\ket{-}$ and $\ket{+}$
are defined as $\frac{1}{\sqrt{2}}(\ket{-1}-\ket{1})$ and $\frac{1}{\sqrt{2}}(\ket{-1}+\ket{1})$,
respectively. The separation between the states $\ket{-}$ and $\ket{+}$
is of the order of $\frac{(\gamma_{e}B)^{2}}{D}$. All three transitions
between the states $\ket{0}$, $\ket{-}$, and $\ket{+}$ are allowed
and have strong dipole moments (Fig. \ref{LACs}(d)). Most importantly,
the orientations of dipole moments of these transitions are orthogonal
to each other. The dipole moment of the transition between the states
$\ket{-}$ and $\ket{+}$ is parallel to the NV axis, the dipole moment
of the transition between the states $\ket{0}$ and $\ket{+}$ is
along the direction of the magnetic field, and the dipole moment of
the transition between the states $\ket{0}$ and $\ket{-}$ is oriented
perpendicular to both these directions. In the coordinate system defined
earlier in this section, $|\mprod{-}{S_{z}}{+}|\approx1$, $|\mprod{0}{S_{x}}{+}|\approx1$,
and $|\mprod{0}{S_{y}}{-}|\approx1$, while all other matrix elements
are close to zero. This can be utilized for vector detection of
magnetic fields \cite{Jingfu_RFsens2023}. The highest sensitivity is achieved for AC fields that are on-resonance with these transitions.

\begin{figure}
\includegraphics[width=8.5cm]{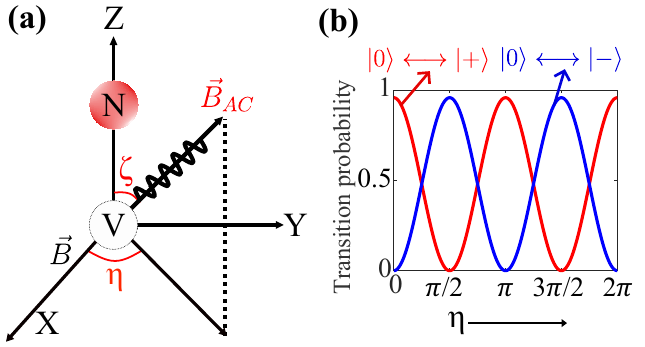} \caption{(a) Coordinate system with static and AC magnetic field orientations.
(b) Transition probabilities of $\ket{0}\protect\longleftrightarrow\ket{+}$
and $\ket{0}\protect\longleftrightarrow\ket{-}$ transitions as a
function of $\eta$ for $\theta=\pi/2$.}
\label{Axes} 
\end{figure}

A linearly polarized AC magnetic field can be expressed as 
\begin{align}
\vec{B}_{AC}= & B_{AC}(\sin\zeta\cos\eta\ \hat{x}+\sin\zeta\sin\eta\ \hat{y}+\cos\zeta\ \hat{z})\nonumber \\
 & \cos(\omega t+\varphi),
\end{align}
where $B_{AC}$, $\zeta$, and $\eta$ are the strength, polar, and
azimuthal angles of the field, respectively. $\omega$ and $\varphi$
are the angular frequency and phase of the field. A schematic representation
of $\vec{B}_{AC}$, and the angles $\zeta$ and $\eta$ are given
in Fig. \ref{Axes}(a). The Hamiltonian corresponding to the interaction
of the AC field with the electron spin of the NV center can be written
as 
\begin{align}
{\cal H}_{AC}=\gamma_{e}\vec{B}_{AC}\cdot\vec{S}.
\end{align}

The purpose of this paper is to determine the parameters $B_{AC}$,
$\zeta$, and $\eta$ of the AC field from experimental measurements.
As shown in Fig. \ref{Axes}(b), the probabilities of the transitions
$\ket{0}\longleftrightarrow\ket{-}$ and $\ket{0}\longleftrightarrow\ket{+}$
have sinusoidal dependence on the angle $\eta$ between the static
field and the transverse component of the AC field. When $\eta=0$
($\eta=\pi/2$), the probability of the transition $\ket{0}\longleftrightarrow\ket{-}$
($\ket{0}\longleftrightarrow\ket{+}$) is close to zero and the probability
of $\ket{0}\longleftrightarrow\ket{+}$ ($\ket{0}\longleftrightarrow\ket{-}$)
is close to one. Therefore, by comparing the experimental amplitudes
or Rabi frequencies of the aforementioned transitions for a fixed
transverse orientation of the static field, we can determine the angle
$\eta$. Once $\eta$ is known, $B_{AC}$ and $\zeta$ can be determined
by comparing the Rabi frequencies of the transition $\ket{-}\longleftrightarrow\ket{+}$
and any one of the two transitions $\ket{0}\longleftrightarrow\ket{-}$
and $\ket{0}\longleftrightarrow\ket{+}$. When the the frequency of
the applied AC field matches the frequency of the transitions, whose
dipole moment is along the direction of the field, it induces transitions,
which may be observed as a drop in the fluorescence emitted by the
NV centers under optical illumination. This effect is known as Optical
Detection of Magnetic Resonance (ODMR) \cite{Bitter_ODMR1949,Suter_ODMRRev2020}.

\section{Proof-of-principle experiments}

The experiments were performed on a home-built confocal microscope
setup equipped with a 532 nm laser for off-resonant excitation of optical
transitions, and radio-frequency (RF) and MW electronic circuits
for resonant excitation of electron spin transitions. A CVD grown
single crystal diamond with a nitrogen concentration of 800 ppb was
used. There were hundreds of NV centers within the excitation volume
of the confocal microscope. The AC magnetic fields, whose vector detection
we were trying to achieve, was generated by a current through
a 25 $\mu$m copper wire attached to the diamond surface. This work
required precise orientation of the static magnetic field with respect
to the NV axis and it was achieved by using a permanent magnet attached
to two rotational stages whose axes were orthogonal to each other
and crossed at the site of the diamond crystal. NV centers of all
four possible orientations that were within the confocal spot were
excited simultaneously by the laser and contributed to the total detected
fluorescence. However, in the ODMR experiments, the frequency of the
applied field selects only one specific orientation.

\begin{figure}
\includegraphics[width=8.5cm]{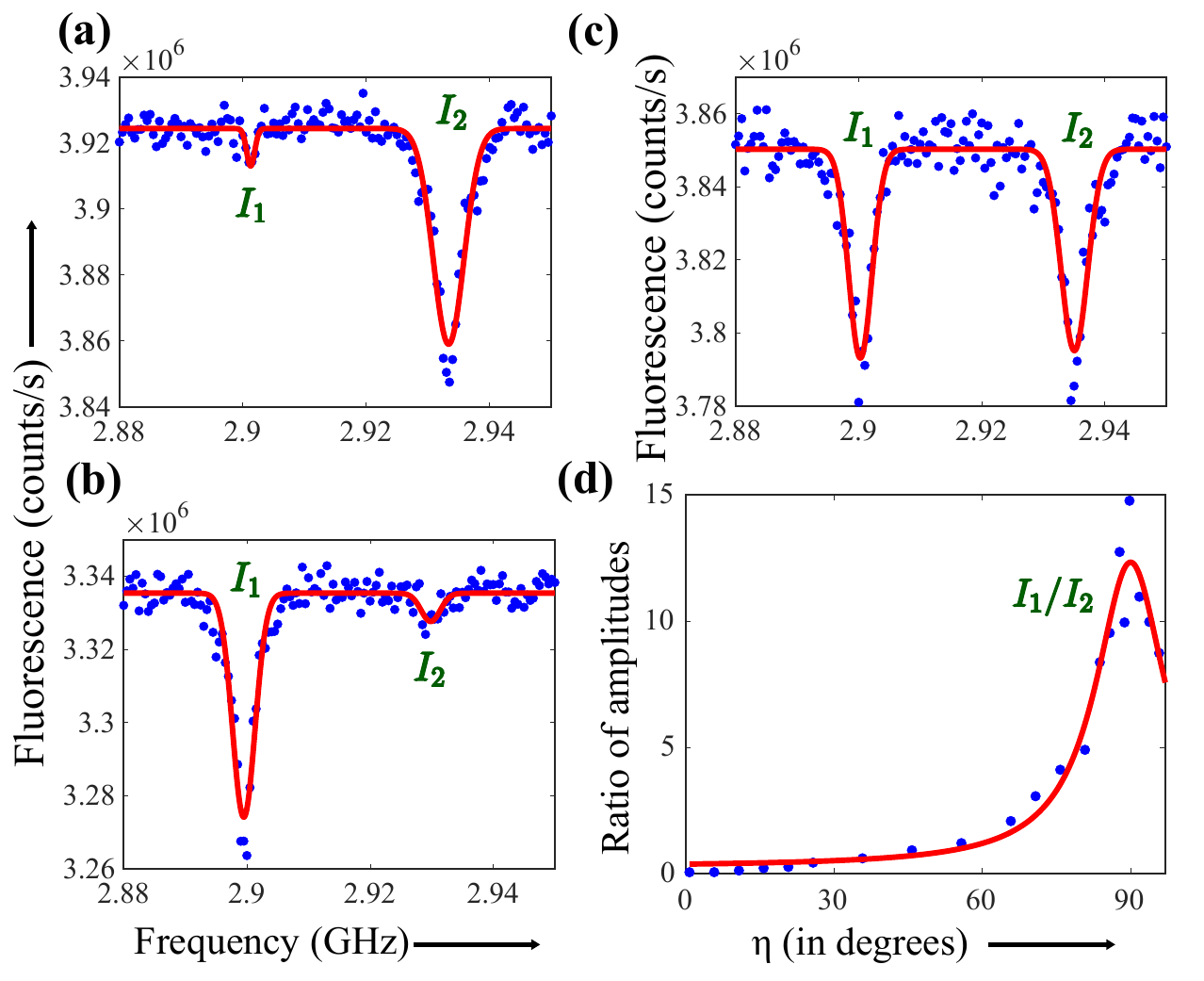} \caption{ODMR spectra for different orientations of the static magnetic field
in the plane perpendicular to the NV axis, (a) $\eta=0$, (b) $\eta=\pi/2$,
(c) $\eta=\pi/4$. Blue dots represent experimental data and red lines
are Gaussian fits of these data. (d) Ratio of amplitudes, $I_{1}/I_{2}$
as a function of $\eta$. Blue dots represent the ratio of amplitudes
determined from the ODMR spectra and the red line represents a fit of these data
to the expression $a\frac{b}{(\eta-\eta_{0})^{2}+b^{2}}+c$, where
$\eta_{0}=90^{\circ}\,(\pm0.9).$}
\label{ODMRspec}
\end{figure}

As described in the previous section, we determine $\eta$, the angle
between the transverse component of the AC field and the static magnetic
field. Specifically, Fig. \ref{LACs}(d) illustrates that when the
static magnetic field is oriented perpendicular to the NV axis, the
transition between the states $\ket{0}$ and $\ket{+}$ can only be
excited by a component of the AC field that is parallel to the static
magnetic field. Likewise, the transition between the states $\ket{0}$
and $\ket{-}$ can be excited by a component of the AC field that
is perpendicular to both static field and NV axis. In our experimental
setup, the direction of the applied AC field is fixed, but we can
change the direction of the static magnetic field. We recorded ODMR
spectra by applying a static magnetic field of strength 10.7 mT at
different orientations in the plane perpendicular to the NV axis.
Fig. \ref{ODMRspec} summarizes the results. When $\eta=0$, the ODMR
consists of only one peak at 2932 MHz corresponding to the $\ket{0}\longleftrightarrow\ket{+}$
transition. The other peak at 2900 MHz corresponding to the transition
$\ket{0}\longleftrightarrow\ket{-}$ has negligible intensity. Conversely,
when $\eta=\pi/2$, the peak at 2932 MHz is of negligible intensity
compared to the peak at 2900 MHz. When $\eta=\pi/4$, as shown in
Fig. \ref{ODMRspec}(c), both peaks are of equal intensity. In other
words, when the static magnetic field is aligned with the transverse
component of the AC field, the transition $\ket{0}\longleftrightarrow\ket{-}$
has minimum intensity and the transition $\ket{0}\longleftrightarrow\ket{+}$
has maximum intensity. So, we can determine the direction of the transverse
component of the AC field by simply rotating the static magnetic field
in the plane perpendicular to the NV axis and look for the direction
at which the transition $\ket{0}\longleftrightarrow\ket{-}$ has minimum
intensity and the transition $\ket{0}\longleftrightarrow\ket{+}$
has maximum intensity. The ratio of intensities of these two transitions
as a function of $\eta$ is given in Fig. \ref{ODMRspec}(d). The
maximum of this ratio corresponds to $\eta=\pi/2$. From this, we
are able to determine the direction of the transverse component of the
AC field with an uncertainty $\pm0.9^{\circ}$. Alternatively, one
can also determine this at a given magnetic field orientation in the
transverse plane by comparing the Rabi frequencies of the two transitions.

\begin{figure}
\centering \includegraphics[width=9cm]{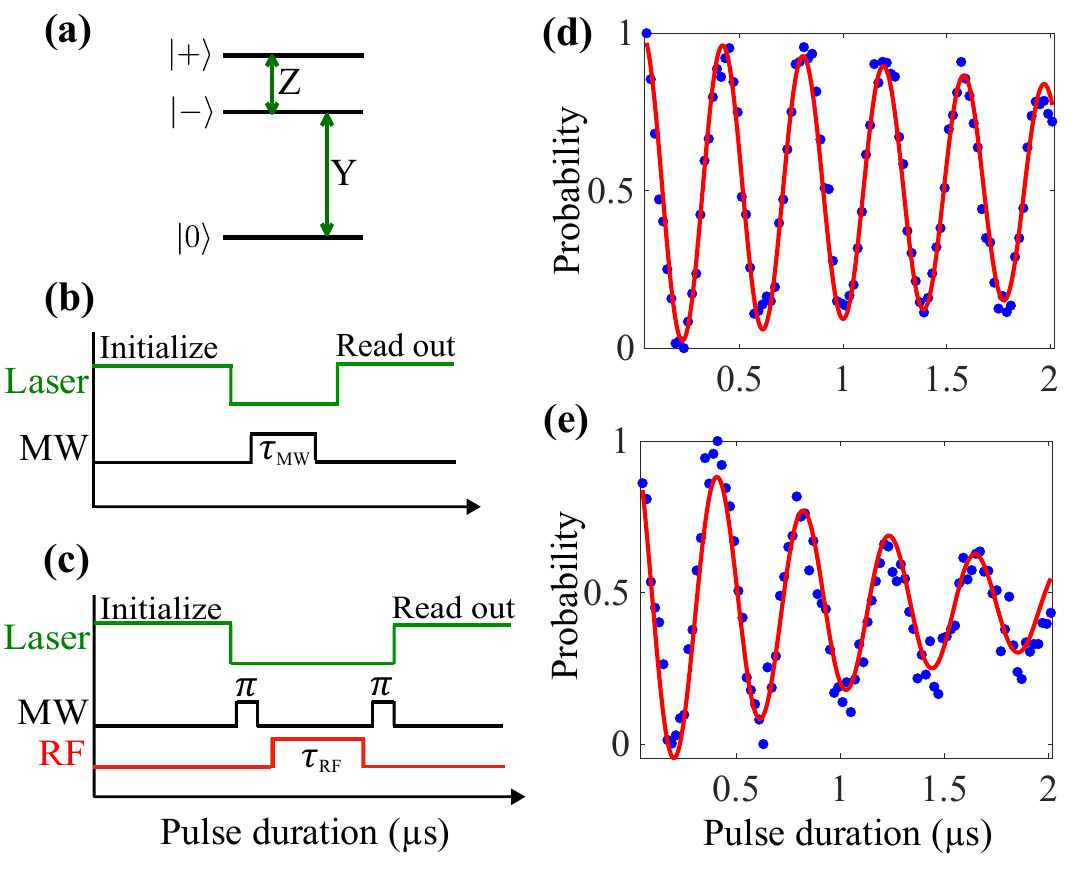}

\caption{Rabi oscillations and the pulse sequences. (a) Electron spin energy
level diagram displaying the transitions whose Rabi oscillations are
measured. (b) and (c) represent the pulse sequences, and (d) and (e)
are the resultant Rabi oscillations of the transitions $\ket{0}\protect\longleftrightarrow\ket{-}$
and $\ket{-}\protect\longleftrightarrow\ket{+}$ respectively. MW
and RF in (b) and (c) refer to the frequencies of the transitions
$\ket{0}\protect\longleftrightarrow\ket{-}$ and $\ket{-}\protect\longleftrightarrow\ket{+}$,
which are 2900 MHz and 32 MHz respectively. In (d) and (e), blue dots
represent experimental data and red lines represent fits of this data
to the expression $a\,\cos(2\pi\nu t)\exp(-t/T_{R})+c$. $\nu=2.57\,(\pm0.01)$
and $2.42\,(\pm0.02)$ MHz, and $T_{R}=5\,(\pm1.7)$ and $1.4\,(\pm0.3)$
$\mu$s for (d) and (e) respectively.}
\label{Rabi}
\end{figure}

As mentioned in the last section, $B_{AC}$, the strength of the applied
AC field and $\zeta$, the angle between the NV axis and the direction
of the AC field can be determined from the measured Rabi frequencies
of the transitions, $\ket{0}\longleftrightarrow\ket{-}$ and $\ket{-}\longleftrightarrow\ket{+}$
for a fixed value of $\eta$. The Rabi oscillations of the transition
$\ket{0}\longleftrightarrow\ket{-}$ can be straightforwardly measured
by using the pulse sequence given in Fig. \ref{Rabi}(b) and the result
is shown in Fig. \ref{Rabi}(d).

Laser illumination doesn't create population difference between the
states $\ket{-}$ and $\ket{+}$ and also these states cannot be distinguished
directly from the intensity of the fluorescence emitted by the NV
centers. So, to measure the Rabi oscillations of the transition $\ket{-}\longleftrightarrow\ket{+}$,
the pulse sequence given in Fig. \ref{Rabi}(c) is used. First, the
state $\ket{0}$ is populated by a laser pulse and this population
is transferred to the $\ket{-}$ state by a microwave $\pi$-pulse.
This creates population difference between the states $\ket{-}$ and
$\ket{+}$. For the actual Rabi experiment, the transition $\ket{-}\longleftrightarrow\ket{+}$
is driven on-resonance by an RF pulse with a frequency of 32 MHz and
variable duration. The resulting dynamics are measured by transferring
the population from $\ket{-}$ to $\ket{0}$ by the second microwave
$\pi$-pulse and reading it out by applying the second laser pulse.
The corresponding Rabi oscillations are shown in Fig. \ref{Rabi}(e).

The expressions relating the Rabi frequencies and the corresponding
transition amplitudes can be written as

\begin{align}
\sqrt{2}\frac{\gamma_{e}}{2\pi}B_{MW}\sin\zeta\cos\eta\ |\mprod{0}{S_{x}}{+}|=R_{0+},\\
\sqrt{2}\frac{\gamma_{e}}{2\pi}B_{MW}\sin\zeta\sin\eta\ |\mprod{0}{S_{y}}{-}|=R_{0-},\\
\sqrt{2}\frac{\gamma_{e}}{2\pi}B_{RF}\cos\zeta\ |\mprod{-}{S_{z}}{+}|=R_{-+}.\label{eq:Rabi}
\end{align}
Here, $B_{MW}$and $B_{RF}$ represent the amplitudes of the AC field
at frequencies 2900 MHz and 32 MHz respectively. $R_{0+}=3.15\ (\pm0.01)$
MHz, $R_{0-}=2.57\ (\pm0.01)$ MHz and $R_{-+}=2.42\ (\pm0.02)$ MHz
are the measured Rabi frequencies of the transitions $\ket{0}\longleftrightarrow\ket{+}$,
$\ket{0}\longleftrightarrow\ket{-}$ and $\ket{-}\longleftrightarrow\ket{+}$
respectively. The ratio $\frac{B_{RF}}{B_{MW}}=0.23$ has been determined
from the measured power levels. By substituting all the extracted
parameters and solving for $B_{MW}$, $B_{RF}$, $\zeta$ and $\eta$,
we obtain them as $B_{MW}=2.85\ (\pm 0.01)$ G, $B_{RF}=0.66\ (\pm 0.01)$ G, $\zeta=21.6^{\circ}\ (\pm 0.2)$
and $\eta=38.8^{\circ} \ (\pm 0.1)$.

The sensitivity of the magnetic field measurement can be calculated
by using the formula \cite{Lukin_Mag2008NatPhy,Harneit_Sens2011PRL,Jacques_Sens2011PRB}
\begin{align}
\eta_{B}=\delta B\sqrt{nT},
\end{align}
where $\delta B$ is the uncertainty in the measurement of the AC
magnetic field, $n$ is the total number of experiments, and $T$
is the sensing time in each experiment. From the fit of the Rabi oscillations,
we estimate $\eta_{B}\approx1\ \mu T/\sqrt{Hz}$, which is similar
to the previously reported sensitivity values \cite{DuVectorMW,Cappellaro_VecAC2021}.

\section{Discussion and conclusion}

\begin{figure}
\includegraphics[width=9cm]{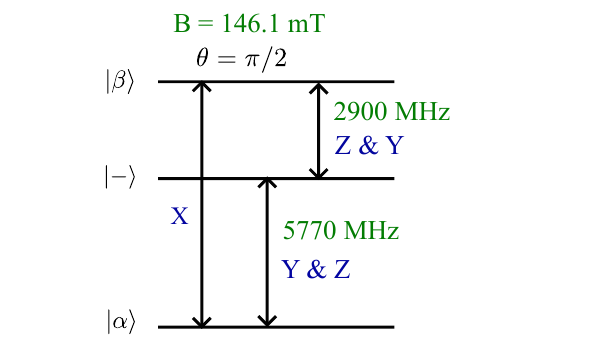} \caption{Energy level diagram of the electron spin in the NV center for $B=146.1$
mT and $\theta=\pi/2$. Here, $\ket{\alpha}=0.82\ket{0}-0.58\ket{+}$
and $\ket{\beta}=0.82\ket{+}+0.58\ket{0}$. The labels X, Y and Z
represent the orientation of the dipole moment of the corresponding
transitions.}
\label{Fig4} 
\end{figure}

The experimental demonstration of the method presented in the last
section uses an AC field of two different frequencies for the vector
detection. However, when vector detection of an arbitrary AC field
is performed, it may not always be possible to tune its frequency.
The proposed method can be used for vector detection of an AC field
of single frequency, but it requires application of a static magnetic
field of two different strengths; one of them is greater than 100
mT, which is beyond the field strength that our current experimental
setup can produce at the site of diamond crystal. Nevertheless, we now discuss the procedure to perform vector detection at a single frequency
- as a specific example, we choose a frequency of 2900 MHz. The angle
$\eta$ or the direction of the transverse component of the AC field
can be determined by applying a static magnetic field of strength
10.6 mT in the plane perpendicular to the NV axis. The frequency of
the transition $\ket{0}\longleftrightarrow\ket{-}$ at this field
is 2900 MHz. The orientation of the static field at which this transition
has the lowest intensity is the direction of the transverse component
of the AC field. To determine the magnitude $B_{AC}$ and the angle
$\zeta$, first we can measure the Rabi frequency of the transition
$\ket{0}\longleftrightarrow\ket{-}$ in the static field of 10.6 mT
oriented in the transverse plane with $\eta=\pi/2$. Next, a field
of 146.1 mT can be applied in the transverse plane, with $\eta=0$.
This field mixes the states $\ket{0}$ and $\ket{+}$ and the new
eigenstates are $\ket{\alpha}$, $\ket{-}$, and $\ket{\beta}$, where
$\ket{\alpha}=0.82\ket{0}-0.58\ket{+}$ and $\ket{\beta}=0.82\ket{+}+0.58\ket{0}$.
The state $\ket{-}$ doesn't mix with the other states, as it is an
eigenstate of both, the $S_{z}^{2}$ and $S_{x}$ operators. The energy
level diagram for this field orientation is given in Fig. \ref{Fig4}.
The frequency of the transition $\ket{-}\longleftrightarrow\ket{\beta}$
is 2900 MHz and it can be excited by the Y- and Z-components of the
AC field. Since we already know the direction of the transverse component
of the AC field, the Y-component can be made equal to zero by orienting
the static magnetic field along this direction. Note that X-axis is
defined as the direction of the static field. So, for $\eta=0$, the
transition $\ket{-}\longleftrightarrow\ket{\beta}$ can only be excited
by the Z-component of the AC field. By measuring the Rabi frequency of
this transition and using it in combination with the Rabi frequency
of the transition $\ket{0}\longleftrightarrow\ket{-}$ at the field
of 10.6 mT, we can determine the parameters, $B_{AC}$ and $\zeta$.
An important point to note here is that the mixing between the states
$\ket{0}$ and $\ket{+}$ can decrease the polarization of the electron
spin under optical pumping and that in turn can decrease the ODMR
contrast.

The aforementioned procedure is intended for vector detection of
AC fields of frequency greater than 2870 MHz as this is the lowest
frequency for the transitions $\ket{0}\longleftrightarrow\ket{-}$
and $\ket{0}\longleftrightarrow\ket{+}$. Now, we discuss the protocol
for vector detection of fields in the frequency range 1 - 2870 MHz.
The frequency of the transition $\ket{-}\longleftrightarrow\ket{\beta}$
is in the frequency range 1 - 2870 MHz for static fields of strength
0 - 145 mT oriented perpendicular to the NV axis. This transition
can be excited by only the Z-component or by Z- and Y-components of
the applied AC fields depending upon the strength and orientation
of the static field in the transverse plane of the NV center. When
a static field of strength 0 - 100 mT is applied parallel to the NV
axis, the frequency of the transition $\ket{0}\longleftrightarrow\ket{-1}$
is in the range 1 - 2870 MHz and it can only be excited by the transverse
components of the AC field. So, from the Rabi frequencies of the transitions
$\ket{-}\longleftrightarrow\ket{\beta}$ and $\ket{0}\longleftrightarrow\ket{-1}$
at appropriate static magnetic fields, we can determine the strength
of the AC field ($B_{AC}$) and the angle between the NV axis and
the AC field ($\zeta$). The Rabi frequency of the transition $\ket{-}\longleftrightarrow\ket{\beta}$ is proportional to $|\cos{\zeta}\ \mprod{-}{S_{z}}{\beta}+\sin{\zeta} \sin{\eta}\ \mprod{-}{S_{y}}{\beta}|$, which is minimum at $\eta=0$ and maximum at $\eta=\pi/2$ for a fixed value of $\zeta$. So, by measuring the Rabi frequency of this transition for different orientations of static field in the transverse plane, we can determine $\eta$, the azimuthal angle of the AC field. However, the amplitude of variation in these Rabi frequencies is proportional to the matrix element $|\mprod{-}{S_{y}}{\beta}|$, which is very small for static fields of strength less than 10 mT. Hence, experimental determination of $\eta$ for AC fields of frequency less than 30 MHz may be difficult.

In conclusion, we have proposed and experimentally demonstrated a
method for vector detection of AC fields by using NV centers having single
orientation. The method is undemanding relative to the convectional technique and is applicable for wide frequency
range. Though the method is demonstrated by using an ensemble of NV
centers with a single orientation, it is equally applicable for an
ensemble consisting only of a single NV center. Since this method
does not require NV centers of multiple orientations, it can be used
for vector imaging of AC fields with nano-scale resolution. In particular, this method is suitable for vector AC field imaging by using single NV centers of diamond nanopillar probes attached to atomic force microscope \cite{Maletinsky_NVAFM2012,Maletinsky_MWsens2015njp}. This work will also be useful for optimal control of quantum registers based on multiple dipolar coupled NV centers having different orientations in diamond. In such registers, it is not possible to align static magnetic field with the NV axes of all the centers. Designing time-optimal quantum gates to the electron and $^{13}$C nuclear spin qubits of misaligned NV centers requires accurate knowledge of the internal as well as the RF and MW control field Hamiltonians.  The vector AC field detection scheme demonstrated in this work is useful for precise determination of the control field Hamiltonian.

\section{Acknowledgements}

A.R. and R.K.K. acknowledge support from Department of Science \& Technology - Science \& Engineering Research Board (DST-SERB), India through grant
no. SRG/2020/000765. S.D. thanks Indian Institute of Technology, Madras,
India for the seed funding. S.D. acknowledges the financial support by the Mphasis F1 Foundation given to the Centre for Quantum Information, Communication,and Computing (CQuICC).

\bibliography{bibNV_LAC}
 
\end{document}